# Diffusive exchange of trace elements between alkaline melts: implications for element fractionation and timescale estimations during magma mixing

*Diego González-García[1], Maurizio Petrelli[1], Harald Behrens[2], Francesco Vetere[1], Lennart A. Fischer[2], Daniele Morgavi[1], Diego Perugini[1]*

[1] *Dipartimento di Fisica e Geologia, Università degli Studi di Perugia, Piazza Università, 06123 Perugia, Italy.*

[2] *Institut für Mineralogie, Leibniz Universität Hannover, Callinstrasse, 3, 30167 Hannover, Germany.*

Corresponding author**: Diego González-García**

e-mail: diego.gonzalez@studenti.unipg.it

**Abstract**

The diffusive exchange of 30 trace elements (Cs, Rb, Ba, Sr, Co, Y, La, Ce, Pr, Nd, Sm, Eu, Gd, Tb, Dy, Ho, Er, Tm, Yb, Lu, Ta, V, Cr, Pb, Th, U, Zr, Hf, Sn and Nb) during the interaction of natural mafic and silicic alkaline melts was experimentally studied at conditions relevant to shallow magmatic systems. In detail, a set of 12 diffusion couple experiments have been performed between natural shoshonitic and rhyolitic melts from the Vulcano Island (Aeolian archipelago, Italy) at a temperature of 1200 °C, pressures from 50 to 500 MPa, and water contents ranging from nominally dry to ca. 2 wt. %. Concentration-distance profiles, measured by Laser Ablation ICP-MS, highlight different behaviours, and trace elements were divided into two groups: (1) elements with normal diffusion profiles (13 elements, mainly low field strength and transition elements), and (2) elements showing uphill diffusion (17 elements including Y, Zr, Nb, Pb and rare earth elements, except Eu). For the elements showing normal diffusion profiles, chemical diffusion coefficients were estimated using a concentration-dependent evaluation method, and values are given at four intermediate compositions ($SiO_2$ equal to 58, 62, 66 and 70 wt. %, respectively). A general coupling of diffusion coefficients to silica diffusivity is observed, and variations in systematics are observed between mafic and silicic compositions. Results show that water plays a decisive role on diffusive rates in the studied conditions, producing an enhancement between 0.4 and 0.7 log units per 1 wt.% of added $H_2O$. Particularly notable is the behaviour of the trivalent-only REEs (La to Nd and Gd to Lu), with strong uphill diffusion minima, diminishing from light to heavy REEs. Modelling of REE profiles by a modified effective binary diffusion model indicates that activity gradients induced by the $SiO_2$ concentration contrast are responsible for their development, inducing a transient partitioning of REEs towards the shoshonitic melt. These results indicate that diffusive fractionation of trace elements is possible during magma mixing events, especially in the more silicic melts, and that the presence of water in such events can lead to enhanced chemical diffusive mixing efficiency, affecting also the estimation of mixing to eruption timescales.

## 1. Introduction

Mass transport by diffusion is involved in several magmatic processes, such as crystal nucleation and growth or dissolution (e.g. Liang, 2003), mass exchange during the mixing of magmas (e.g. Morgavi et al., 2016, Laeger et al., 2016), or through the interaction between magmas and the host rocks or xenoliths (e.g. Watson, 1982). Being a time-dependent process, diffusion is often used to estimate the timescales of geological processes. Examples are the study of diffusion profiles in zoned crystals (e.g. Demouchy et al., 2006; Chakraborty, 2008; Costa et al., 2008, Faak et al., 2014) and melt inclusions (e.g. Ferguson et al., 2016), as well as the use of rocks showing evidence of magma mixing as volcanic chronometers (e.g. Perugini et al., 2015, Rossi et al., 2017). For a successful application of petrological techniques based on the diffusion process, an accurate knowledge of the diffusive behaviour for the investigated chemical elements is mandatory.

In the last decades, a large number of studies provided a great wealth of data about diffusivities for major and trace elements in a wide range of silicate melt compositions (for a detailed summary of diffusion data in silicate melts, see Zhang et al., 2010, and references therein). However, the currently available dataset still lacks diffusivity data for many elements, especially for natural melt compositions. For example, the first data of rare earth elements (REE) diffusion coefficients in basalt were just recently provided by Holycross and Watson (2016). Effects of dissolved water on diffusion in magmas are poorly constrained, and investigations on hydrous magmas are often focused on silicic compositions (Watson, 1979;

Watson, 1981; Harrison and Watson, 1983, 1984; Rapp and Watson, 1986; Baker, 1991; Watson, 1994; Baker et al., 2002; Mungall et al., 1999). Furthermore, diffusion studies using natural rock compositions, that are potentially more relevant for geological applications, are scarce in the literature (e.g. Baker, 1990; Lundstrom, 2006).

Recently, González-García et al. (2017) reported results on the diffusive exchange of major elements in shoshonite-rhyolite couples. Here we use the same set of experiments to investigate the trace element diffusion in natural silicate melt compositions at physical conditions relevant for subvolcanic magmatic systems. Our primary objectives are: (1) to investigate the diffusive behaviours of trace elements in the presence of strong compositional gradients, linking the results to major element diffusivity; and (2) to assess the influence of water on trace element diffusivities. The implications of this study for element fractionation during mixing of magmas and timescale estimations of related geological processes are discussed.

## 2. Materials and methods

### 2.1 Starting materials

Two different volcanic products with contrasting compositions from the Vulcano island (Aeolian Archipelago, Italy) were selected as end-member compositions for the experiments. The mafic composition is a shoshonite (Vetere et al., 2007; Davì et al., 2009) from the Vulcanello lava platform, dating from a long-lasting eruption series in the period between the 2nd and 16th centuries A.D. (Keller, 1980; Arrighi et al., 2006; De Astis et al., 1997). The evolved composition is a high-K rhyolite sampled in the 1739 A.D. Pietre Cotte obsidian lava flow (Keller, 1980; Frazzetta et al., 1983). These two compositions are the most contrasted ones erupted by the Vulcano system in historical times, and several pieces in the erupted products evidence complex magma mixing events (Vetere et al., 2015b; Clocchiatti et al., 1994). Their major and trace element compositions are summarized in Table 1. Fig. 1 shows the TAS diagram (Le Bas et al., 1986) of both end-members, and chondrite-normalized compositions and REE patterns are presented in Fig. 2. End-member compositions follow similar patterns, with negative anomalies in Nb-Ta and Ti, which are distinctive of subduction magmas (e.g. Pearce, 1982). In the REE diagram, both end-members show similar light REE abundances, with a small Eu anomaly in the shoshonite and a pronounced one in the rhyolite. The Sr and Eu negative spikes in the rhyolite suggest that crystallization of plagioclase was involved in its genesis. The abundance of heavy REEs of the end-members diverge, with the rhyolite progressively being more enriched towards the heavier elements. Concentration contrasts are significant, especially for Ba, Sr, Cr, V and Eu.

## 2.2 Experimental setup

Natural rock samples were cleaned with distilled water and crushed to fine powder. Subsequently, these powders were melted in air at 1600°C for 4 h (Nabertherm® HT 04.17) at atmospheric pressure inside a Pt crucible and quenched to produce a glass (for details, please refer to Vetere et al., 2015a). The glass was then crushed, and the process was repeated, to ensure a homogenous glass. Finally, the produced glass was crushed again before the preparation of the experimental capsules.

Hydrous glasses were produced at high pressure and temperature using $Au_{80}Pd_{20}$ alloy tubes with a diameter of 5 mm and a length of 40 mm. The capsules were filled with rock powder and, when necessary, distilled water, in several steps (the procedure is described in detail in Vetere et al., 2014). Glasses of both end-members were produced as nominally dry (ND, i.e. no added water) and adding water up to nominal $H_2O$ concentrations of 1 wt.% and 2 wt.%. After loading, capsules were welded shut, and then glasses were synthetized at 300 MPa and 1200 °C in an Internally Heated Pressure Vessel (IHPV; Berndt et al., 2002) for 24h. The samples were quenched isobarically after switching off the furnace. Such a procedure gives an initial cooling rate of ca. 200 °C/min. For quality control of the glass synthesis procedure, bulk analyses were carried out in chips from the two ends of the produced glass cylinders, in order to ensure homogeneity. Karl Fischer titration (KFT) was used to measure the water content, and iron oxidation state ($Fe^{2+}/Fe^{3+}$ ratio) was characterized by a colorimetric wet-chemical analysis (Schuessler et al., 2008). After the synthesis, the glass samples were cut to obtain cylinders with a length of ca. 5-6 mm. Finally, each cylinder was finely polished on one side.

Diffusion experiments were performed using the diffusion couple method. In detail, couples of rhyolite and shoshonite glass cylinders with the same nominal water content were juxtaposed by their polished side and placed in an $Au_{80}$-$Pd_{20}$ capsule. The capsules were hermetically closed by welding. Then, capsules were compressed in a cold seal pressure vessel (CSPV) to check for tightness and to observe the adaptation of the capsule to sample shape.

The experiments were run in a vertically operating IHPV at a temperature of 1200°C and pressures of 50, 100 and 300 MPa (plus one additional experiment at 500 MPa). The denser shoshonitic glass was placed at the bottom to avoid gravitational instability. The IHPV was pressurized with argon and the experiments were run at the intrinsic oxygen fugacity of the vessel which was estimated by Berndt et al. (2002) to be close to the $MnO$-$Mn_3O_4$ buffer (NNO+3.7) at water saturated conditions. Since the water content of the melts was below the saturation limit, oxygen fugacities were lower in our experiments. In our rhyolitic and shoshonitic glasses produced in the IHPV, $Fe^{2+}/\Sigma Fe$ ratios were, respectively, in the order of 0.7 and 0.85 in dry glasses; and 0.6 and 0.7 in hydrous (2 wt.% $H_2O$) glasses (González-García et al., 2017). This would indicate an $fO_2$ of roughly -7.5, slightly above the NNO buffer for the dry melts (Kress and Carmichael, 1991). Experimental time ranged between 1 h and 4 h (plus one additional zero-time experiment at 2 wt. $H_2O$ and 300 MPa). Heating ramps of 30, 50 and 20 °C/min from ambient temperature to 100, 1150 and 1200 °C, respectively, were used. After each run, a rapid quench device (Berndt et al., 2002) was activated, allowing the capsule to fall into the cold part of the vessel and cool quickly. A total of 12 experiments (including the zero-time experiment) were run successfully and utilized for the present study (Table 3).

*2.3 Analytical procedures*

A double polished section was prepared for each experiment, with thickness of ca. 200 μm. Concentrations of $H_2O$, major elements and trace elements were obtained by Fourier Transform infrared spectroscopy (FTIR), electron microprobe (EPMA) and laser ablation-inductively coupled plasma-mass spectroscopy (LA-ICP-MS), respectively. In each case, chemical concentrations were determined along a linear profile centred at the interface between the two end-members and extending to 1-1.5 mm to each side (resulting in a total profile length of ca. 2-3 mm).

Water concentration by FTIR was obtained using a Bruker IFS88 spectrometer coupled to an IR-Scope II microscope at the Institut für Mineralogie, Leibniz Universität Hannover, Germany (for details please refer to González-García et al., 2017).

Major element oxide concentration profiles ($SiO_2$, $TiO_2$, $Al_2O_3$, FeO, MgO, MnO, CaO, $Na_2O$, $K_2O$) were obtained with a Cameca SX-100 electron microprobe at the Institut für Mineralogie, Leibniz Universität Hannover. Operating conditions were an accelerating voltage of 15 kV, a beam current of 4 nA and a defocused beam diameter of 10 μm, to minimize alkali loss in the glass analysis. Spacing of analyses in the central region of the profiles was kept as small as possible (typically 12-13 μm) and was widened to 30 μm at the end. Precision and accuracy were determined by measuring VG-568 (rhyolite) and VG-2 (basalt) reference glasses (Jarosewich et al., 1980; Helz et al., 2014). Obtained analytical errors vary from 1% for $SiO_2$ to 10% for minor oxides. In addition, to check for reproducibility, and for possible convection effects a second profile was acquired parallel to the first one but with a shift of ca. 1 mm towards the rim (for major element data in these experiments, the reader is referred to González-García et al., 2017).

Concentration profiles of 30 trace elements (Cs, Rb, Ba Sr, Co, Y, La, Ce, Pr, Nd, Sm, Eu, Gd, Tb, Dy, Ho, Er, Tm, Yb, Lu, Ta, V, Cr, Pb, Th, U, Zr, Hf, Sn and Nb) in the diffusion couple glasses were obtained at the Department of Physics and Geology, University of Perugia. The instrumentation consisted in a Teledyne Photon Machine G2 laser ablation device equipped with the two-volume HelEx 2 cell coupled to a Thermo Fischer Scientific iCAP Q quadrupole mass spectrometer (Petrelli et al., 2016a). A circular 20 µm spot size and 20 µm spot spacing was used in most of the analyses with the exception of experiment P300-H2-4, where 25 microns were used for both spot size and spacing. To avoid interference and keep a suitable resolution, every second spot was offset 20 microns to one side, configuring a zig-zag pattern (Fig 3). Ablation times were 25 seconds per spot, preceded by a 25 second background measurement and followed by 25 seconds of washout. Data reduction was completed with Iolite v3.32 software (Paton et al., 2011). NIST SRM 610 glass (Pearce et al., 1997) was analysed as reference material for the external calibration, and the basaltic glass BCR2G (Wilson, 1997) was used as quality control. The $SiO_2$ profiles obtained by electron microprobe were used as internal standard. Analytical precision is about 12% for concentrations close to 0.1 ppm and better than 5% above 20 ppm; accuracy is always better than 10% (Petrelli et al., 2016b).

## 3. Results

Quenched experimental products are crystal-free glasses and show variable amounts of bubbles. Samples from 300 and 500 MPa experiments are bubble-free while in experiments at 50 and 100 MPa small amounts of bubbles were observed near the contact of the diffusion halves (González-García et al., 2017). Given the small size of the bubbles and their hypothesized composition ($N_2$ and minor amounts of water in the hydrous runs), they are considered to have a negligible effect in the diffusive process. Although some water flux between melts can be present during experimental heating, $H_2O$ profiles produced by FTIR analyses show uniform water content across the couples (González-García et al., 2017), indicating that no net flux of $H_2O$ was present during experiment due to water activity contrasts between the end-members.

Concentration-distance profiles of six major elements ($SiO_2$, $TiO_2$, FeO, MgO, CaO, and $K_2O$) are characterized by the expected asymmetric sigmoidal shapes, extending deeper into the mafic melts due to different diffusivities in melts of different polymerization degree. A particular case is that of $Al_2O_3$, with evident signs of uphill diffusion in the rhyolitic side of the couple (but lack of maximum in the shoshonite), an effect that can be explained and predicted by coupled diffusion (e.g. Liang et al., 1996) by using the same diffusion equations. The diffusive behaviour of major element is discussed in detail by González-García et al. (2017). Here we focus on the description and the modelling of trace element diffusivities.

Examples for trace element concentration-distance profiles are reported in Figs. 4 and 5. Based on these figures, elements can be separated in two distinct groups: (I) elements showing normal diffusion profiles (Fig. 4), and (II) elements showing the effects of uphill diffusion (Fig. 5). For the complete dataset, see Electronic Annex 1.

Elements with a normal behaviour (group I) are Rb, Cs, Sr, Ba, Co, V, Cr, Eu, Th, U, Sn, Ta and Hf. These elements are characterized by sigmoidal shape profiles with variable degrees of asymmetry, caused by variation in base composition along the profile (Fig. 4). In binary variation diagrams, elements with normal diffusive behaviour show linear to slightly non-linear trends (S-shape), in which the deviation from linearity correlates with the difference in diffusion rates between the two involved elements. Sr has a virtually identical behaviour to that of Ba, and hence they show a very good linear relation while V, Cr or Eu, characterized by higher diffusivity contrasts, produce more prominent deviations from the linear trend when compared to Ba.

The second group of elements (group II) displays different extent of uphill diffusion, i.e., diffusion against the concentration gradient of the diffusing element (Fig. 5). Uphill diffusion has often been observed for major element diffusion in multicomponent silicate melts (e.g. Guo and Zhang, 2016), and few studies have described the same effect for trace elements (Zhang et al., 1989; van der Laan et al., 1990; Richter, 1993; Lesher, 1994). The elements included in this group are Y, Zr, Nb, Pb and, most notably, all trivalent-only REEs (La to Sm and Gd to Lu; Fig 5). The exception among REEs is Eu, which shows a normal diffusion profile. Uphill diffusion is evident in the profiles by concentration minima and maxima near the original interface between the two halves of the couples. We have observed a prominent effect in La, Ce, Pr, Nd and Sm (the light REEs), for which a very deep concentration minimum is identified, reaching concentrations well below that of the end members. This minimum is usually accompanied by a wide and less prominent maximum in the shoshonitic side. It is worth noting that group II elements are always elements with low starting concentration gradients (Table 1), with more prominent minima in the REEs with more similar initial starting concentrations in both melts (e.g. for Nd). Binary variation elements of group II elements always result in highly asymmetric trends, with compositions extending much higher and lower than the starting end-members compositions. The deep minimum correlates to Ba concentration of ca. 69 wt.% $SiO_2$ and the weak maximum near 54 wt% $SiO_2$ is present. No appreciable differences in distribution are observed when considering experiments at different pressure, time and water contents.

Since the heating times are significant in relation to the experiment times, an approximate value of effective ramp-up times ($t_{eff}$) was estimated in the 300 MPa, 2 wt.% $H_2O$ experiment set by comparing the concentration-distance profiles of the zero-time experiment to that of 1 hour and 4 hour experiments (Zhang and Behrens, 2000). Results indicate effective heating times to be in the order of 80 to 120 seconds, short enough to not significantly affect the experimental results.

A common test to prove possible undesirable effects (e.g. convection) in experiments with different duration, is to plot the concentration versus distance normalized by the square root of time. Applying this methodology to the REE profiles, datasets from experiments P300-H2-4 (4 h) and P300-H2-1 (1 h) collapse to the same curve (Fig. 6). This is evidence that the process producing this kind of profiles is controlled only by diffusion.

## 4. Discussion

### 4.1. Determination of diffusion coefficients for group I elements

Chemical diffusion coefficients were determined for all 13 group I trace elements displaying normal diffusion profiles. Given the evident asymmetry of the profiles, a simple, constant diffusivity model cannot be applied. Hence, diffusivities were obtained along the profiles using a concentration-dependent diffusivity approach (Sauer and Freise, 1962), which is a modification of the original Boltzmann-Matano method (Boltzmann, 1894; Matano, 1933). These methods have been used when diffusivities vary as a function of the concentration of the diffusing element itself (e.g. Nowak and Behrens, 1997), and also when diffusivities depend on the concentrations of other elements (Baker, 1989). The procedure involves: (1) smoothing the analytical profiles by a single polynomial function (for examples, refer to Electronic Annex 2), (2) normalization of the compositional range and (3) applying the solution for one dimension, molar volume independent diffusion as follows:

$$D(x) = \frac{1}{-2t(\partial c/\partial x)}\left[(1 - c(x)) \int_x^\infty c\,dx + c(x) \int_{-\infty}^x \big(1 - c(x)\big)dx\right] \qquad (1)$$

where $x$ is the distance along the profile (in meters); $t$ is the experimental time (in seconds); $D(x)$ is the diffusion coefficient at $x$ (in m²/s): $c(x)$ is the normalized concentration of the diffusing component [where $c(-\infty) = 1$ and $c(+\infty) = 0$]. The origin of coordinates (x = 0) is chosen to be the initial position of the interface. However, it should be noted that the Sauer-Freise method is not sensitive to the origin of the coordinate system. The complete procedure was implemented in a Python (Oliphant, 2007) script.

The modified Boltzmann-Matano method of Sauer and Freise (1962) produces a continuous dataset of diffusion coefficients along the profiles (Fig. 7), in a way that each diffusion coefficient can be related to a particular chemical composition. Along this profile, four diffusion coefficients are extracted at intermediate compositions corresponding to 20%, 40%, 60% and 80% of the major element profile corresponding to a latite (Lt$_{58}$), trachytes (Tr$_{62}$ and Tr$_{66}$) and a rhyolite (Rh$_{70}$), where subscripts denote the SiO$_2$ concentration in wt%. Major element compositions of these four intermediate melts are summarized in Table 2.

Table 3 summarizes all chemical diffusion coefficients ($D$) calculated for the 13 elements belonging to group I and characterized by normal diffusion profiles. As a general rule, diffusivities are noticeably higher in the most mafic intermediate composition (Lt$_{58}$) than in the most silicic (Rh$_{70}$), although some exceptions exist. The most evident case is that of Ba and Sr, whose diffusivities are insensitive to the silica content of the melt. This is already observable in Fig. 4, where Ba and Sr show symmetric profiles. The application of the Sauer-Freise method allows to confirm the lack of variation of Ba and Sr with bulk melt composition. For the remaining elements, a smooth increase of diffusivities from rhyolite to shoshonite is observed, with varying slopes (Fig. 8). Relations between diffusivity and SiO$_2$ content can be fit by a linear equation. Melt composition has a relatively low effect in LFSE and TE, but the effect noticeably increases in HFS elements (Th, U, Sn and Ta), where the diffusivity increase from Rh$_{70}$ to Lt$_{58}$ varies between 0.6 and 0.9 log units.

Rough systematic variations of trace element diffusivities with field strength ($z^2/r$, $z$ being the ionic charge and $r$ the ionic radius) have been observed by previous works (e.g. Mungall, 2002; Behrens and Hahn, 2009). Among our data, systematic trends are only evident in the silicic compositions, where diffusivities increase from Cs to Rb-Ba, and then follow a negative trend until Ta-Cr. HFSE elements do not present well defined trend, and high diffusivities occur for U, Th and Sn. However, this systematic trend gradually disappears towards progressively more mafic melts. In the $Lt_{58}$ composition, systematic correlations are no longer observable, although U, Th and Sn still show high diffusivities (Fig. S1 in the Electronic Annex 2). Trace element diffusivities are mostly clustered around the value of $SiO_2$ diffusivity measured in the same experiments (González-García et al., 2017), and particularly in the more mafic compositions where the variation in diffusivities is small around that value. These observations suggest a strong control of trace element diffusivity by $SiO_2$ mobility (i.e., timescale of relaxation processes in the melt). Individual element migration processes are thus at least partially coupled with the Si-O dynamics (breaking and re-formation of Si-O bonds). With increasing silica content, LFSE elements become relatively more mobile than silica while Ta, V, Cr become relatively less mobile. The trends for the LFSEs towards higher silica content can be explained by less dense packing of structural units in polymerized networks. Data for Cs show that this constraint does not apply to very large cations, consistent with the model of Mungall (2002).

As reported in Table 3, the most determining factor for the diffusion rates is the presence of water. Fig. 8 shows how diffusion coefficients are related to water content in the system. The relation between the water content and *log D* can be fit by a linear equation, in the form *log D* = *a*\**w* + *b,* in which *a* is the slope and *w* is the water concentration in weight percent. We also find that in the 50-500 MPa range, pressure does not have a significant impact on the diffusion coefficients and, as consequence, we can be confident in the reliability of this linear equation to the pressure range mentioned above. Table 4 summarizes the calculated equation parameters and their relative errors.

Moreover, the effect of water is not uniform for all measured elements. This is evident in Fig. 9 where the slope in $log\,D$ vs $H_2O$ plot, ($a$ parameter, fig. 10 and Table 4) is plotted against elements sorted by increasing field strength. The effect of water results to be highest for LFSE (Rb, Ba, Sr), where the value of $a$ ranges between 0.6 and 0.7 log units per 1 wt.% increase of $H_2O$ concentration. U and Th also display very high values and comparable to that of LFSE. On the contrary, the effect of $H_2O$ has a minimum in the transition elements Ta and Cr (0.4 to 0.5 log units per 1 wt. $H_2O$). Eu, Hf, Sn and V show intermediate values, in agreement with the lower water effect on the diffusion of Eu and Cr already reported by Koepke and Behrens (2001) for andesitic melts. Finally, no noticeable difference of water effect on diffusivities can be observed between mafic and felsic melts (Fig. 11).

Among the rare earth elements, Eu is the only one to show a normal diffusion profile, allowing the calculation of chemical diffusion coefficients. Contrary to the remaining REEs, Eu can be present in two different oxidations states ($Eu^{2+}$ and $Eu^{3+}$). It has been found that $Eu^{2+}$ has a similar diffusive behaviour to $Sr^{2+}$ (Behrens and Hahn, 2009), an element that shows a normal diffusion profile in our experiments. Reversely, $Eu^{3+}$ is expected to have a behaviour similar to Gd and Sm, which in our dataset are dominated by uphill diffusion. In our experiments, measured Eu diffusivities are between 0.25 and 0.5 log units lower than that of Sr in the $Rh_{70}$ melt composition. Hence, although the diffusion behaviour resembles that of divalent cations, a large fraction of $Eu^{3+}$ is still present in the melts. However, quantification of the redox state of Eu is not possible from our experiments since diffusion coefficients for trivalent REEs could not be determined. Large $Eu^{3+}/Eu_{tot}$ ratios in our experiments are supported by Behrens and Hahn (2009) from experiments carried out in the same IHPV working with variably hydrous melts, where $Eu^{3+}/Eu_{tot}$ ratios were found to be between 0.6 to 0.9, and similar values were obtained experimentally by Burnham et al., 2015. However, these values are not enough to explain the observed shift in behaviour from uphill to normal, which is very likely the consequence of a change in initial conditions, i.e. starting concentration contrast (Eu is over 7 times more abundant in the shoshonitic end-member than in the rhyolitic one) and direction of diffusive flux.

A similar reasoning is valid for Sn. It has one of the highest diffusivities, most prominently shown in the shoshonitic side, where diffusion coefficients are higher than -11 log units, resulting a high compositional variation from $Rh_{70}$ to $Lt_{58}$. This is comparable to data in Behrens and Hahn (2009) for trachyte and phonolite with similar oxygen fugacity and water contents, although at significantly higher alkali content. Sn is effectively the element with highest diffusivity in the present study. However, its water dependence is not particularly high, being comparable to that of Co and Eu. On the other hand, Sn diffusivities are one order of magnitude higher than the tetravalent cation $Ti^{4+}$ (González-García et al., 2017); and the diffusivities in $Rh_{70}$ are over one order of magnitude higher than the values obtained by Yang et al. (2016) by means of cassiterite dissolution experiments in haplogranite melt with low oxygen fugacity. This evidence suggests that $Sn^{2+}$ is contributing significantly to Sn diffusivities.

### 4.2. Viscosity-diffusion systematics

Given the indications of a systematic variation of diffusivities, especially in the more silicic compositions, an attempt was made to relate the measured diffusion coefficients with the correlation schemes proposed by Fanara et al. (2017) based on viscosity-diffusivity relations. In their work, a convergence towards Eyring diffusivity was observed for melts with low viscosities, with a progressively higher decoupling from Eyring diffusivity at higher melt viscosities. This trend was first suggested by Dingwell (1990) and is not uncommon in the literature (e.g. Behrens and Stelling, 2011). Observed power laws of cations of different ionic charge differ, thus resulting in a different degree of decoupling. This results in a low interelemental diffusivity variation at low viscosities and a more differential behaviour at higher viscosities.

The Eyring equation is defined as follows:

$$D = \frac{k_B * T}{\lambda * \eta} \quad (2)$$

where $k_B$ is the Boltzmann constant, $\lambda$ is the interatomic jump distance and $\eta$ is the viscosity of the melt. This equation provides a way to relate diffusion of network former elements to melt viscosity (e.g. Zhang, 2010). For the calculation of the Eyring diffusivities, a jump distance of 0.4 nm is assumed in this work, a value similar to atomic spacing in silicates.

A comparison between diffusivities measured in this work and melt viscosities is shown in Figure 10, where elements are separated by ionic charge. Melt viscosities were calculated using the model described by Giordano et al. (2008) and are listed in Table 3. Experimental diffusivities are always higher than calculated Eyring diffusivity, and as previously outlined, diffusivities in melts with lower viscosities (most mafic and water-rich) tend to approach more to Eyring diffusivity that that of higher viscosity (most felsic and water-poor). However, we note that no single trend can be observed for all melts, and instead a different behaviour is observed for melts with different water content. This is especially evident in the divalent cations. Due to the restricted range of our melt viscosities (log($\eta$) < 4, with $\eta$ in Pa s) compared and the involved uncertainties (0.40 log units in viscosity: Giordano et al. 2008), the trends from monovalent, divalent and trivalent cations do not differ strongly, and fall below the equations provided by Fanara et al. (2017) for divalent and trivalent cations.

Figure 10 includes data for the tetravalent cations U, Th, Hf and Sn (although Sn, as previously noted, probably consists in a mixture of $Sn^{2+}$ and $Sn^{4+}$), a group not considered by Fanara et al. (2017). A high interelemental dispersion is observed in this group, with decoupling between particular elements (i.e., no single trend can be observed for all of them). On the other hand, a single trend appears to be valid for both dry and hydrous melts. Results show that tetravalent cations diverge from Eyring diffusivity at lower viscosities, following the behaviour of monovalent, divalent and trivalent cations. In this plot, Sn stands out from U, Th and Hf, which can be interpreted as further evidence of a significant presence of $Sn^{2+}$ in the melt.

## 4.3. Treatment of uphill diffusion

A method to model trace element uphill diffusion is the modified effective binary diffusion approach proposed by Zhang (1993) in which activity gradients, instead of concentration gradients, are used to describe the diffusive flux (Zhang, 1993; Richter, 1993; Lesher, 1994). In this model, the basic approach is described by

$$J_i = -D_i \frac{\nabla a_i}{\gamma_i} \qquad (3)$$

where $J_i$ is the diffusive flux of component $i$, $\boldsymbol{D}_i$ is referred as the intrinsic diffusivity, $a_i$ is the activity and $\gamma_i$ the activity coefficient. In this approach, diffusing components would tend to achieve a constant activity (i.e., chemical potential), allowing them to diffuse against their own concentration gradient if the activity coefficient varies with composition. An alternative way to visualize this effect would be to use melt-melt partition coefficients: uphill diffusion could thus be viewed as an element trying to achieve transient partition equilibrium between two structurally different melts. The concept of transient equilibrium during interdiffusion of complex silicate melts was originally introduced by Watson (1982). In addition, it has been observed that in experiments with two immiscible melts, alkalis are enriched in the more polymerized melts (Watson and Jurewicz, 1984), and many others (most notably, HFSE) are enriched in less polymerized melts (Zhang, 1993; Watson, 1976). To achieve this transient equilibrium (Watson, 1982), some components may need to move against their own concentration gradients. The equilibrium concentration $C_e$ to which elements would tend to accommodate is defined as (Zhang, 1993):

$$C_e = \frac{C_{-i} + C_{+i}}{2} + C_f\, erf\, \frac{x}{2\sqrt{D_b t}} \qquad (4)$$

where $C_{-i}$ and $C_{+i}$ are the initial concentrations of element $i$ in the two end-members; $D_b$ is the diffusivity that can be assimilated to $SiO_2$ or $SiO_2 + Al_2O_3$ (network forming species) effective binary diffusivity in the system; and $C_f$ is half the difference between the equilibrium concentrations in both sides of the couple. This parameter is related to the two-liquid partition coefficient $K$ as follows (Zhang, 1993):

$$K = \frac{C_{+e}}{C_{-e}} = \frac{(C_{+e} + C_{-e})/2 + C_f}{(C_{+e} + C_{-e})/2 - C_f} \qquad (5)$$

The concentration-distance profiles resulting from diffusion of a component in this context can be modelled by

$$\frac{\partial C_i}{\partial t} = \frac{\partial}{\partial x}\left(D_i C_e\, \frac{\partial\left(C_i / C_e\right)}{\partial x}\right) \qquad (6)$$

As given by equations (4), (5) and (6), in order to quantitatively model the uphill diffusion phenomena observed here by the modified EBD approach, a precise knowledge of three variables is mandatory: (1) diffusivity of $SiO_2$ or $SiO_2 + Al_2O_3$ (network formers), (2) intrinsic EBD coefficients (i.e., self-diffusivities) of trace elements, and (3) the value of $C_f$, which defines the equilibrium concentration and is related to the melt-melt partition coefficient $K_{sho/rhy}$. Although we have insufficient data of two of these variables ($C_f$ and $D_i$), it is possible to use the modified EBD model to obtain information on the causes and evolution of uphill diffusion in the shoshonite-rhyolite system. We have implemented the modified EBD approach described above (Zhang, 1993) in a script in Python programming language, using a forward finite differences model (Press et al., 1986). For simplicity, $SiO_2$ effective binary diffusivities were modelled by a step function with a different value for shoshonite and rhyolite halves of the couple. Values were extrapolated for 53 wt.% $SiO_2$ and 73 wt.% $SiO_2$ from diffusivities obtained in the same experiments (González-García et al., 2017).

Very scarce data is available on the self-diffusion of trace elements in literature. In the present modelling, we have used the value Nd self-diffusion in basalt-rhyolite diffusion couple experiments provided by Lesher (1994) as reference, which is in the range of 2.5 to 3.3 times the diffusivity of $SiO_2$. Finally, the value of $C_f$ was adjusted in the finite difference model to obtain a good visual fit of the profiles.

Results show a good agreement between the model and the experiments (Fig. 11) and confirm the partitioning of REEs, Y, Zr, Nb and Pb towards less polymerized melts. In other words, activities are significantly higher in silicic melts than in mafic melts, resulting in a partition flux towards the shoshonitic side of the couples. Table 5 summarizes the values used to model the uphill diffusion profiles. The model produces an inflection in the interface position of all elements, which is a consequence of assuming one value of $D_{SiO_2}$ for each end-member, and thus producing a discontinuous first derivative $dC/dx$. Uphill diffusion arises as a consequence of the interplay between initial concentration difference and the variation of activity coefficients. The shape of the diffusion profile is the result of the interplay between the end-member starting concentrations and the transient partitioning of the element between shoshonitic and rhyolitic melts, as already observed by Watson (1982) and van der Laan et al., (1994); the asymmetry is the consequence of the major element diffusivity contrast in the melts (here represented by $SiO_2$ diffusivity) and the intrinsic diffusivity of the trace element. Therefore, the higher diffusivities present in the most mafic melt produce the wider and lower maximum observed in the shoshonitic side of the profiles. In the LREE, the small concentration contrast and the high value of the partition coefficient result in very prominent uphill diffusion. On the contrary, the higher compositional contrast in the HREE between both halves and the lower value of $K_{Sho/Rhy}$ result in progressively milder uphill diffusion (Fig. S2 in the Electronic

Annex 2). The model suggests that the value of $K_{sho/rhy}$ is around 3 for LREE, and diminishes to roughly 1.6 for the heaviest REES (Tm, Yb and Lu). Similarly, best fits were obtained using intrinsic diffusivity factors ($D_i$) varying from 3.5 (La, Ce and Pr) to 2 (Tm, Yb and Lu). Y, Zr and Nb can be modelled with parameters similar to that of the HREE elements. Finally, Pb shows a more longitudinally extended profile than other group II elements, and in consequence, it can be only modelled assuming a much higher intrinsic diffusivity factor of $\sim$10. The same parameters are valid for both nominally dry and hydrous experiments through our experimental pressure range. The modelled partition data are similar in magnitude with previously published experimental partitioning data in immiscible rhyolitic and basaltic liquids, where partition coefficients of REE were found to be in the range between 3.7 and 4.3 (Watson, 1976; Ryerson and Hess, 1978), and also to that in natural immiscible melts observed in lunar melt inclusions (Shearer et al., 2001).

The modified EBD model is also capable to reproduce the behaviour of group I elements for which intrinsic diffusivities and melt-melt partitioning data are known (fig. 11 and Table 5). Eu can be fit reasonably well by assuming parameters similar to that of Sm and Gd, and this is also the case for Sr, which can be modelled assuming an intrinsic diffusivity of 10x$D_{SiO2}$ and $K_{sho/rhy}$ of 2.5 (Lehser, 1994). These results indicate that the primary cause behind the normal behaviour of europium is very likely a change in initial conditions (that is, initial concentration gradient and direction of diffusive flux) with respect to the trivalent REEs, and that the presence of a fraction of $Eu^{2+}$ is not necessarily related. This is expected from previous studies in multicomponent systems, in which uphill diffusion in a component can disappear by changing the direction of diffusion in the compositional space, and hence indicating that this feature does not only depend on activity gradients (e.g. Chakraborty et al., 1995).

## 5. Implications

Results presented here support an elemental fractionation process between the fastest (the LFS elements) and slowest diffusivities (V, Hf, Nb). Given that the diffusivity contrast is lower in the mafic end-member, the fractionation trends would be more prominent towards the more evolved component. A suitable fractionation indicator could thus be the Ba/V ratio. Ba diffusion is around one order of magnitude faster than V in hydrous $Rh_{70}$ melt, but only 0.3 orders of magnitude in hydrous $Lt_{58}$. As a result, the Ba/V ratio vs. distance plot (Fig. 12) displays a maximum in the rhyolitic side while keeping the ratio approximately constant in the shoshonite. The maximum fractionation is Ba/V = 25, markedly higher than the ratios measured in the end-members at great distances from the interface.

Although trace element uphill diffusion is a striking feature in the experiments presented throughout this work, its potential to produce relevant observable features in natural environments remains unknown. According to literature studies on simple melts composed by only few components, uphill diffusion in major elements is a temporary feature that disappears once strong concentration gradients relax (e.g. Liang, 2010), and similar conclusions can be extracted for trace elements (Lesher, 1994). As a consequence, its visibility would be limited to systems in which the diffusive flux has been frozen in an early stage, as it is the case for pre-eruptive magma mixing processes. A definitive sign of the development of uphill diffusion in natural systems would be the observation of strongly nonlinear trends in binary diagrams. Uphill diffusion would, in addition, produce strong concentration variations on short length scales. Magma mixing experiments performed in dynamic conditions support this idea. In fact, Perugini et al. (2008) and Perugini et al. (2013) have shown highly nonlinear concentration-distance profiles that could be attributed to the occurrence of transient partition leading to uphill diffusion.

This idea is contrasting to the general assumption that magma mixing tends to the chemical homogenization of the system. While it is true for major element composition and some of the trace elements, here we show that uphill diffusion has the opposite effect. Compositional gradients for elements diffusing uphill would tend to increase in the early phases of the mixing process, even when their end-member concentrations do not differ strongly. This finding puts some doubts on the use of REEs as a proxy to obtain timescales in magma mixing systems, when strong $SiO_2$ gradients are present.

One of the main results presented here is that water is a very important conditioning factor for trace element diffusion, and its effect is not the same for all diffusing cations. Since magmas in nature are rarely dry, especially in subduction environments where water content can reach 6 to 8 wt.% (Wallace, 2005; Plank et al., 2013), this is a fundamental aspect. The concept of diffusion distance can be used to quantify the effect of the addition of water in mixing times. The distance ($x$) to which a certain cation diffuses in a given time ($t$) is given by the expression $x^2 = Dt$. As it is straightforward from this simple equation (assuming constant diffusivity), the diffusion distance is proportional to the square root of time. On the other hand, the time to reach a certain diffusion distance will be proportional to the diffusion coefficient. As a consequence, the time to achieve the same degree of diffusive homogenization with melts of different water content depends on the enhancement factor $a$ produced by water. We have seen that for LFS elements, going from 0.3 to 1.9 wt.% increases the diffusivity by 1.2 to 1.4 orders of magnitude, which in turn would decrease the time required to achieve the same degree of homogenization by the same amount. For transition elements, this value would be approximately one order of magnitude. Therefore, it can be concluded that using dry diffusion coefficients to model the mixing of magmas with a certain amount of water could lead to overestimations in timescales by factor up to 25.

It should be noted that these considerations can be applied only to the restricted water concentration range covered in this study (up to 2 wt.% $H_2O$), where we find that the influence of water in $log\ D$ can be described by a linear relation. However, several previous studies with larger water concentration ranges have found that reaching 3-4 wt.% $H_2O$, the linear relation is lost and a plateau is achieved above ~4 wt.% $H_2O$ (e.g., Harrison and Watson, 1983).

## 6. Conclusions

The diffusive exchange of 30 trace elements between shoshonitic and rhyolitic melts of natural compositions was measured in diffusion couple experiments by using nominally dry and hydrous conditions. Results show a complex scenario, where the strong compositional gradient induces non-ideal behaviours in many trace elements, mainly REEs. A group of elements comprising LFS, TE and some HFSE display normal behaviours, and thus chemical diffusion coefficients were calculated. The major factor governing their diffusivities is the water content dissolved in the melt, which greatly enhances their diffusivity. Diffusion of LFS (Ba, Sr, Cs, Rb) are notably enhanced (0.7 log units increase per 1 wt.% of added $H_2O$), while TE (Ta, Cr) show a moderate enhancement (0.4-0.5 log units per 1 wt. $H_2O$). Diffusivities increase smoothly from mafic ($Tr_{58}$) to silicic ($Rh_{70}$) compositions, with the exception of Ba and Sr, whose diffusivities are insensitive to the anhydrous melt composition. The systematic variation of diffusivity in silicic melts is in contrast to that of mafic melts: in the $Rh_{70}$ composition, a clear variation of diffusivity with field strength is observed. However, this correlation is gradually lost towards the mafic compositions, where diffusivities are roughly uniform. In the whole compositional spectrum, Sn, U and Th show high diffusivities.

A second group of elements, mostly REE, show evident signs of non-ideal behaviour. Deep uphill diffusion minima and maxima are observed at the interface region, and this effect is most pronounced in the light REEs. This behaviour is attributed to the influence of the $SiO_2$ gradient, resulting in activity gradients that interfere with normal diffusion. The modified EBD model of Zhang, 1993 is capable of reproducing this behaviour in REEs, Y, Zr, Nb and Pb, and suggest a transient melt-melt partitioning towards the shoshonitic, less polymerized melt.

In summary, our results show that, in a natural system with strong compositional gradients, the diffusive exchange is complex and it is dominated by the large $SiO_2$ compositional gradients. This is in agreement with the observations reported in González-García et al. (2017), where strong diffusive coupling between major elements was observed. As a consequence, it should be noted that diffusion coefficients obtained in melts with constant bulk compositions are not appropriate to model complex mixing systems.

These new results have implications for the estimation of homogenization rates and timescales in magma mixing events. The influence of dissolved water content in the mixing melts is an essential factor to consider, resulting in a notable acceleration of the mass exchange rate, and therefore, the relative timescale estimations. Fractionation of trace elements is a possible process occurring mainly between fast-diffusing LFSEs and slow-diffusing TEs and HFSEs. Uphill diffusing phenomena in natural environments is still unclear, but it could potentially result in small-scale heterogeneities and highly nonlinear variation diagrams. These features would be best observed in magma mixing events that are frozen by eruption in an early stage, where compositional gradients are still high.

**Acknowledgements**

This research was funded by the European Research Council Consolidator Grant ERC-2013-COG No. 612776 (CHRONOS project) to D. Perugini and by the MIUR-DAAD joint mobility Project No. 57262582 to F. Vetere and H. Behrens. We wish to acknowledge the assistance received from R. Balzer during experiment run and J. Feige for the careful preparation of experiment sections, both of them at the Insitut für Mineralogie, Leibniz Universtät Hannover; and also A. Zezza for completing the first diffusion experiment. Constructive comments by E.B. Watson, Y. Liang and an anonymous reviewer significantly improved the final version of this manuscript.

**Figure captions**

Fig. 1: End-member and intermediate compositions plotted in a total alkali vs silica diagram (TAS; Le Bas et al., 1986). Data correspond to the average of points where diffusivities were calculated, and error bars represent the standard deviations.

Fig. 2: (A) Chondrite normalized trace element (Thompson, 1982) and (B) rare earth element (Boynton, 1984) diagrams of starting materials.

Fig. 3: Detail of a laser ablation spot track showing the zig-zag pattern used during analysis in experimental polished sections.

Fig. 4: Normalized concentration-distance profile of the 13 group I elements with normal diffusion profiles. Experiment P300-H2-1 (1 h, 2 wt.% nominal $H_2O$ and 300 MPa).

Fig. 5: Concentration-distance profiles of selected REEs displaying variable uphill diffusion (left axis) compared to the $SiO_2$ profiles measured by EPMA (right axis). Data belongs to experiment P300-H2-1 (300 MPa, 1 wt.%, 4h). Error bars representing 2σ are reported below the element symbol, and starting profiles are shown as solid lines.

Fig. 6. Concentration-distance profiles of La, Gd and Lu with distance normalized by the square root of time from experiments P300-H2-4 and P300-H2-1 (200 MPa, 2 wt.% nominal), with durations of 4h and 1h respectively. Error bars representing 2σ are reported below the element symbol. Curves for both experiments are coincident, as expected if only diffusion is playing a role.

Fig. 7: Selected diffusivity vs. $SiO_2$ content diagrams resulting from the Sauer and Freise (1962) calculation method for the 300 MPa experiment set.

Fig. 8: Relation between water content and the diffusivity of Sr and V in the pressure range 50-500 MPa.

Fig. 9: Systematic variation of water influence in diffusion (*a* parameter, the slope of linear fits) given in log units per 1 wt.% added $H_2O$, with elements arranged by increasing ionic field strength ($z^2/r$).

Fig. 10: Relation between elemental diffusivity and calculated melt viscosity (Giordano et al., 2008), with elements separated by ionic charge. The red line represents the Eyring diffusivity, for a jump value of 0.40 nm. Experimental diffusivities fall always above Eyring diffusivity and tend to converge towards it at lower viscosities, with differing systematics for each element group, following the correlation schemes of Fanara et al. (2017).

Fig. 11. Results of the application of the modified EBD model (Zhang, 1993) to uphill diffusion of some REE, Y, Zr, Nb and Pb. Experimental data are from experiment P300-H2-1, and the red line is the predicted concentration modelled from the parameters in Table 5. Error bars representing $2\sigma$ of the analytical data are reported below the element symbol. The modified EBD model is also capable of modelling Sr and Eu (bottom row), elements with normal diffusion profiles with known model parameters. The model produces a kink in the interface position, consequence of assuming one value $SiO_2$ diffusivity.

Fig 12. Fractionation ratio Ba/V represented versus distance (A) and versus $SiO_2$ (B) for 2 wt.% and 300 MPa, plotted for different experimental times.

**Table captions**

Table 1: Major and trace element compositions of end-member glasses. Major elements were acquired by electron microprobe and are normalized to 100%. Trace elements were acquired by LA-ICP.MS. Each value is the average of 10 measurements, and standard deviation ($\sigma$) is given in italics.

Table 2: Major element abundances of the four intermediate compositions where diffusivities are extracted. Compositions correspond to 20%, 40%, 60% and 80% of the compositional range, and are the average of points where diffusivities are extracted. Standard deviation ($\sigma$) is given in italics. Quoted viscosities were calculated with the Giordano et al. (2008) method and correspond to an anhydrous composition.

Table 3: Diffusion coefficients of group I elements (elements showing normal diffusion profiles), calculated by the modified Boltzmann-Matano method of Sauer and Freise (1962)

with their estimated errors. Quoted viscosities were calculated by the Giordano et al. (2008) general viscosity model. Values of estimated errors are given in italics.

Table 4: Parameters of the linear fits relating water content and diffusivity with their corresponding errors. Equations are in the form $log D = a*w + b$, where $a$ is the slope and $b$ the zero ordinate. Estimated errors are quoted in italics.

Table 5. Parameters used to reproduce the uphill diffusion behaviour of Group II elements Y, Zr, Nb, Pb and REE using the modified EBD model (Zhang, 1993). Group I elements Sr and Eu are also included (bottom rows). $D_i$ is the intrinsic diffusivity of the element; $D_{SiO2}$ is the effective binary diffusion coefficient of $SiO_2$; $C_f$ is half the difference between equilibrium concentrations at the end-members; and $K_{Sho/Rhy}$ is the resulting melt-melt partition coefficient.

**Table 1**

|  | Shoshonite | | Rhyolite | |
|---|---|---|---|---|
|  | wt.% | $\sigma$ | wt.% | $\sigma$ |
| $SiO_2$ | 53.34 | *0.77* | 73.20 | *0.67* |
| $TiO_2$ | 0.69 | *0.04* | 0.11 | *0.03* |
| $Al_2O_3$ | 16.42 | *0.15* | 13.84 | *0.31* |
| FeOt | 8.14 | *0.28* | 2.14 | *0.26* |
| MgO | 4.64 | *0.10* | 0.18 | *0.04* |
| MnO | 0.21 | *0.13* | 0.08 | *0.09* |
| CaO | 8.04 | *0.18* | 0.92 | *0.18* |
| $Na_2O$ | 5.46 | *0.21* | 4.22 | *0.20* |
| $K_2O$ | 3.05 | *0.07* | 5.31 | *0.06* |
|  | ppm | $\sigma$ | ppm | $\sigma$ |
| V | 192.4 | *1.4* | 10.11 | *0.21* |
| Cr | 41.4 | *2.8* | 2.99 | *0.17* |
| Co | 26.45 | *0.28* | 1.71 | *0.07* |
| Rb | 124.0 | *1.3* | 275.2 | *2.5* |
| Sr | 1157 | *13* | 74.63 | *0.49* |
| Y | 19.86 | *0.09* | 35.06 | *0.29* |
| Zr | 150.8 | *1.1* | 185.59 | *1.3* |
| Nb | 19.62 | *0.16* | 33.93 | *0.52* |
| Sn | 1.96 | *0.16* | 5.69 | *0.16* |
| Cs | 5.95 | *0.18* | 15.75 | *0.20* |
| Ba | 1035.3 | *16.2* | 76.76 | *0.65* |
| La | 57.12 | *0.83* | 65.21 | *0.44* |
| Ce | 106.7 | *1.8* | 122.8 | *1.0* |
| Pr | 11.23 | *0.22* | 12.35 | *0.11* |
| Nd | 42.09 | *0.85* | 42.33 | *0.44* |
| Sm | 7.70 | *0.18* | 8.03 | *0.10* |
| Eu | 1.67 | *0.03* | 0.22 | *0.01* |
| Gd | 5.69 | *0.13* | 6.32 | *0.11* |
| Tb | 0.71 | *0.01* | 0.93 | *0.01* |
| Dy | 3.97 | *0.07* | 6.06 | *0.06* |
| Ho | 0.73 | *0.01* | 1.22 | *0.02* |
| Er | 2.03 | *0.03* | 3.72 | *0.06* |
| Tm | 0.29 | *0.01* | 0.59 | *0.01* |
| Yb | 1.93 | *0.03* | 4.06 | *0.07* |
| Lu | 0.28 | *0.01* | 0.60 | *0.01* |
| Hf | 3.67 | *0.04* | 6.11 | *0.06* |
| Ta | 1.13 | *0.02* | 2.37 | *0.04* |
| Pb | 13.80 | *1.53* | 22.42 | *0.38* |
| Th | 23.53 | *0.29* | 51.00 | *0.46* |
| U | 4.94 | *0.12* | 15.07 | *0.13* |

**Table 2**

| | Intermediate compositions | | | |
|---|---|---|---|---|
| | Lt$_{58}$ | Tr$_{62}$ | Tr$_{66}$ | Rh$_{70}$ |
| | Latite | Trachyte | Trachyte | Rhyolite |
| SiO$_2$ | 58.15 *(0.58)* | 62.25 *(0.62)* | 66.45 *(0.66)* | 70.02 *(0.91)* |
| TiO$_2$ | 0.61 *(0.03)* | 0.42 *(0.02)* | 0.27 *(0.03)* | 0.15 *(0.04)* |
| Al$_2$O$_3$ | 15.74 *(0.26)* | 14.99 *(0.23)* | 14.04 *(0.32)* | 13.11 *(0.23)* |
| FeO$_t$ | 6.32 *(0.47)* | 5.02 *(0.34)* | 3.98 *(0.58)* | 3.16 *(0.48)* |
| MgO | 3.67 *(0.22)* | 2.79 *(0.23)* | 2.08 *(0.38)* | 1.35 *(0.26)* |
| MnO | 0.08 *(0.07)* | 0.10 *(0.08)* | 0.09 *(0.08)* | 0.09 *(0.04)* |
| CaO | 6.51 *(0.28)* | 5.08 *(0.23)* | 3.60 *(0.30)* | 2.39 *(0.38)* |
| Na$_2$O | 5.20 *(0.55)* | 5.19 *(0.52)* | 4.91 *(0.74)* | 4.85 *(0.50)* |
| K$_2$O | 3.57 *(0.15)* | 4.08 *(0.16)* | 4.55 *(0.22)* | 4.86 *(0.18)* |
| NBO/T | 0.41 | 0.32 | 0.23 | 0.17 |
| log η (Pa s) | 2.56 | 3.03 | 3.49 | 3.87 |

**Table 3**

| Run | P050-H0-4 | P100-H0-4 | P300-H0-1 | P050-H1-4 | P100-H1-4 | P300-H1-4 | P050-H2-4 | P100-H2-4 | P300-H2-1 | P300-H2-4 | P500-H2-4 |
|---|---|---|---|---|---|---|---|---|---|---|---|
| $H_2O$ (wt%) | 0.27 0.03 | 0.35 0.05 | 0.32 0.03 | 1.11 0.08 | 1.02 0.08 | 1.14 0.08 | 1.59 0.10 | 1.88 0.12 | 1.95 0.15 | 1.91 0.14 | 1.64 0.08 |
| P (Mpa) | 50 | 100 | 300 | 50 | 100 | 300 | 50 | 100 | 300 | 300 | 500 |
| Time (s) | 14400 | 14400 | 3600 | 14400 | 14400 | 14400 | 14400 | 14400 | 3600 | 14400 | 14400 |

**Lt58**

| | P050-H0-4 | P100-H0-4 | P300-H0-1 | P050-H1-4 | P100-H1-4 | P300-H1-4 | P050-H2-4 | P100-H2-4 | P300-H2-1 | P300-H2-4 | P500-H2-4 |
|---|---|---|---|---|---|---|---|---|---|---|---|
| log $\eta$ ($\eta$ in Pa s) | 2.59 | 2.51 | 2.54 | 2.01 | 2.05 | 2.00 | 1.81 | 1.72 | 1.69 | 1.71 | 1.80 |
| log D (D in m²/s) Cs | -12.39 0.27 | -12.37 0.16 | -12.20 0.14 | -11.86 0.12 | -11.97 0.11 | -11.92 0.16 | -11.65 0.09 | -11.58 0.12 | -11.46 0.09 | -11.48 0.07 | -11.92 0.08 |
| Rb | -12.33 0.24 | -12.27 0.17 | -12.28 0.17 | -11.76 0.21 | -11.84 0.11 | -11.66 0.10 | -11.58 0.12 | -11.34 0.08 | -11.22 0.09 | -11.29 0.08 | -11.79 0.08 |
| Ba | -12.86 0.18 | -12.37 0.15 | -12.53 0.23 | -12.06 0.13 | -11.87 0.14 | -11.89 0.13 | -11.88 0.09 | -11.53 0.14 | -11.40 0.11 | -11.54 0.08 | -11.77 0.09 |
| Sr | -12.64 0.13 | -12.61 0.16 | -12.24 0.27 | -12.06 0.11 | -12.05 0.11 | -11.77 0.08 | -11.79 0.09 | -11.64 0.13 | -11.43 0.10 | -11.46 0.07 | -11.68 0.10 |
| Co | -12.49 0.24 | -12.69 0.16 | -12.11 0.14 | -12.08 0.16 | -11.92 0.10 | -11.72 0.13 | -11.68 0.14 | -11.53 0.12 | -11.42 0.07 | -11.35 0.10 | -11.80 0.12 |
| Eu | -12.62 0.24 | -12.58 0.16 | -12.36 0.22 | -12.10 0.14 | -12.22 0.14 | -11.89 0.10 | -11.59 0.12 | -11.72 0.19 | -11.57 0.15 | -11.56 0.07 | -11.90 0.11 |
| Ta | -12.18 0.30 | -12.43 0.18 | -11.95 0.21 | -11.81 0.23 | -11.85 0.16 | -11.71 0.15 | -11.62 0.17 | -11.55 0.14 | -11.39 0.16 | -11.52 0.25 | -11.63 0.18 |
| V | -12.65 0.12 | -12.36 0.17 | -12.45 0.20 | -12.13 0.12 | -12.13 0.10 | -11.94 0.12 | -11.88 0.12 | -11.78 0.09 | -11.55 0.08 | -11.60 0.09 | -11.75 0.12 |
| Cr | -12.47 0.20 | -12.45 0.16 | -12.31 0.22 | -12.05 0.14 | -12.05 0.17 | -11.87 0.12 | -11.80 0.19 | -11.56 0.12 | -11.64 0.19 | -11.75 0.15 | -11.87 0.13 |
| Th | -12.16 0.24 | -12.27 0.16 | -12.28 0.19 | -11.81 0.17 | -11.76 0.13 | -11.69 0.17 | -11.61 0.12 | -11.28 0.11 | -11.20 0.10 | -11.23 0.14 | -11.64 0.18 |
| U | -12.41 0.19 | -12.32 0.18 | -12.13 0.21 | -11.72 0.13 | -11.64 0.14 | -11.60 0.18 | -11.41 0.16 | -11.29 0.15 | -11.12 0.14 | -11.19 0.21 | -11.56 0.13 |
| Hf | -12.64 0.30 | -12.61 0.23 | -12.41 0.24 | -12.22 0.19 | -12.03 0.13 | -12.20 0.21 | -11.96 0.20 | -11.68 0.11 | -11.71 0.22 | -11.70 0.13 | -11.96 0.18 |
| Sn | -11.80 0.17 | -11.69 0.15 | -11.61 0.23 | -10.99 0.13 | -11.18 0.08 | -11.34 0.16 | -10.94 0.13 | -10.77 0.15 | -10.82 0.14 | -10.85 0.11 | -10.91 0.15 |

**Tr62**

| | P050-H0-4 | P100-H0-4 | P300-H0-1 | P050-H1-4 | P100-H1-4 | P300-H1-4 | P050-H2-4 | P100-H2-4 | P300-H2-1 | P300-H2-4 | P500-H2-4 |
|---|---|---|---|---|---|---|---|---|---|---|---|
| log $\eta$ ($\eta$ in Pa s) | 3.06 | 2.97 | 3.00 | 2.42 | 2.47 | 2.41 | 2.20 | 2.09 | 2.07 | 2.08 | 2.18 |
| log D (D in m²/s) Cs | -12.68 0.27 | -12.64 0.16 | -12.45 0.14 | -12.14 0.12 | -12.12 0.11 | -12.06 0.16 | -11.88 0.09 | -11.66 0.12 | -11.67 0.09 | -11.58 0.07 | -12.07 0.08 |
| Rb | -12.55 0.24 | -12.54 0.17 | -12.50 0.17 | -11.97 0.21 | -11.98 0.11 | -11.85 0.10 | -11.73 0.12 | -11.34 0.08 | -11.36 0.09 | -11.40 0.08 | -11.88 0.08 |
| Ba | -12.90 0.18 | -12.59 0.15 | -12.70 0.23 | -12.13 0.13 | -11.92 0.14 | -11.92 0.14 | -11.92 0.09 | -11.63 0.14 | -11.44 0.11 | -11.58 0.08 | -11.89 0.09 |
| Sr | -12.83 0.13 | -12.70 0.16 | -12.48 0.27 | -12.12 0.11 | -12.16 0.11 | -11.87 0.08 | -11.85 0.09 | -11.68 0.13 | -11.56 0.10 | -11.56 0.07 | -11.77 0.10 |
| Co | -12.73 0.19 | -12.76 0.16 | -12.32 0.14 | -12.20 0.16 | -12.06 0.10 | -11.87 0.13 | -11.92 0.14 | -11.66 0.12 | -11.50 0.10 | -11.55 0.07 | -11.94 0.12 |
| Eu | -12.84 0.24 | -12.65 0.16 | -12.44 0.22 | -12.25 0.14 | -12.32 0.14 | -11.99 0.10 | -11.93 0.12 | -11.91 0.19 | -11.73 0.15 | -11.63 0.07 | -12.06 0.11 |
| Ta | -12.57 0.30 | -12.66 0.18 | -12.20 0.21 | -12.12 0.23 | -12.17 0.16 | -11.92 0.15 | -11.97 0.17 | -11.66 0.14 | -11.66 0.16 | -11.73 0.25 | -12.00 0.18 |
| V | -12.79 0.12 | -12.73 0.17 | -12.58 0.20 | -12.32 0.12 | -12.30 0.10 | -12.15 0.12 | -12.12 0.12 | -11.98 0.09 | -11.81 0.09 | -11.81 0.09 | -12.03 0.12 |
| Cr | -12.61 0.20 | -12.60 0.16 | -12.46 0.22 | -12.20 0.14 | -12.23 0.17 | -12.00 0.12 | -11.97 0.19 | -11.78 0.12 | -11.71 0.19 | -11.84 0.15 | -12.13 0.13 |
| Th | -12.49 0.24 | -12.47 0.16 | -12.55 0.19 | -11.93 0.17 | -12.03 0.13 | -11.99 0.17 | -11.78 0.12 | -11.53 0.10 | -11.53 0.10 | -11.54 0.14 | -11.99 0.18 |
| U | -12.80 0.19 | -12.73 0.18 | -12.39 0.21 | -12.11 0.13 | -11.97 0.14 | -11.92 0.18 | -11.82 0.16 | -11.55 0.15 | -11.50 0.14 | -11.50 0.11 | -11.80 0.13 |
| Hf | -12.81 0.30 | -12.69 0.23 | -12.48 0.24 | -12.47 0.19 | -12.31 0.13 | -12.37 0.21 | -12.11 0.20 | -11.78 0.11 | -11.94 0.22 | -11.87 0.13 | -12.03 0.18 |
| Sn | -11.93 0.17 | -11.99 0.15 | -11.97 0.23 | -11.20 0.13 | -11.37 0.08 | -11.62 0.16 | -11.06 0.13 | -11.10 0.15 | -11.10 0.14 | -11.18 0.11 | -11.11 0.15 |

**Tr66**

| | P050-H0-4 | P100-H0-4 | P300-H0-1 | P050-H1-4 | P100-H1-4 | P300-H1-4 | P050-H2-4 | P100-H2-4 | P300-H2-1 | P300-H2-4 | P500-H2-4 |
|---|---|---|---|---|---|---|---|---|---|---|---|
| log $\eta$ ($\eta$ in Pa s) | 3.53 | 3.43 | 3.32 | 2.84 | 2.89 | 2.82 | 2.59 | 2.47 | 2.44 | 2.46 | 2.57 |
| log D (D in m²/s) Cs | -12.84 0.27 | -12.76 0.16 | -12.57 0.14 | -12.29 0.12 | -12.19 0.11 | -12.14 0.16 | -11.97 0.09 | -11.71 0.12 | -11.80 0.09 | -11.63 0.07 | -12.20 0.08 |
| Rb | -12.63 0.24 | -12.67 0.17 | -12.59 0.17 | -12.12 0.21 | -12.05 0.11 | -11.92 0.10 | -11.83 0.12 | -11.54 0.08 | -11.47 0.09 | -11.45 0.08 | -11.94 0.08 |
| Ba | -12.93 0.18 | -12.59 0.15 | -12.71 0.23 | -12.11 0.13 | -11.93 0.14 | -11.93 0.14 | -11.90 0.09 | -11.62 0.14 | -11.45 0.11 | -11.56 0.08 | -11.93 0.09 |
| Sr | -12.98 0.13 | -12.65 0.16 | -12.56 0.27 | -12.11 0.11 | -12.19 0.11 | -11.91 0.08 | -11.82 0.09 | -11.68 0.13 | -11.59 0.10 | -11.59 0.07 | -11.81 0.10 |
| Co | -12.89 0.19 | -12.81 0.16 | -12.50 0.14 | -12.29 0.16 | -12.15 0.10 | -11.96 0.13 | -12.03 0.14 | -11.73 0.12 | -11.63 0.07 | -11.58 0.10 | -12.04 0.12 |
| Eu | -12.97 0.24 | -12.73 0.16 | -12.52 0.22 | -12.36 0.14 | -12.42 0.14 | -12.12 0.10 | -12.12 0.12 | -12.03 0.19 | -11.81 0.15 | -11.70 0.07 | -12.19 0.11 |
| Ta | -12.83 0.30 | -12.91 0.18 | -12.46 0.21 | -12.37 0.23 | -12.32 0.16 | -12.05 0.15 | -12.16 0.17 | -11.95 0.14 | -11.86 0.25 | -12.00 0.18 | -12.17 0.18 |
| V | -13.00 0.12 | -12.92 0.17 | -12.65 0.20 | -12.45 0.12 | -12.47 0.10 | -12.30 0.12 | -12.32 0.12 | -12.10 0.09 | -11.95 0.08 | -11.95 0.09 | -12.23 0.12 |
| Cr | -12.73 0.20 | -12.74 0.19 | -12.62 0.22 | -12.33 0.14 | -12.35 0.17 | -12.09 0.12 | -12.10 0.19 | -11.89 0.12 | -11.79 0.19 | -11.94 0.15 | -12.32 0.13 |
| Th | -12.91 0.24 | -12.74 0.16 | -12.76 0.19 | -12.32 0.17 | -12.16 0.13 | -12.19 0.17 | -12.12 0.12 | -11.77 0.11 | -11.65 0.10 | -11.64 0.14 | -12.14 0.18 |
| U | -13.00 0.19 | -12.89 0.18 | -12.56 0.21 | -12.32 0.13 | -12.26 0.14 | -12.24 0.18 | -12.03 0.16 | -11.72 0.15 | -11.67 0.14 | -11.71 0.11 | -11.91 0.13 |
| Hf | -12.98 0.30 | -12.74 0.23 | -12.64 0.24 | -12.57 0.19 | -12.48 0.13 | -12.41 0.21 | -12.32 0.20 | -11.93 0.11 | -12.03 0.22 | -11.97 0.13 | -12.04 0.18 |
| Sn | -11.99 0.17 | -12.45 0.15 | -12.22 0.23 | -11.35 0.13 | -11.54 0.08 | -11.75 0.16 | -11.22 0.13 | -11.26 0.15 | -11.31 0.16 | -11.35 0.11 | -11.41 0.15 |

**Rh78**

| | P050-H0-4 | P100-H0-4 | P300-H0-1 | P050-H1-4 | P100-H1-4 | P300-H1-4 | P050-H2-4 | P100-H2-4 | P300-H2-1 | P300-H2-4 | P500-H2-4 |
|---|---|---|---|---|---|---|---|---|---|---|---|
| log $\eta$ ($\eta$ in Pa s) | 3.91 | 3.81 | 3.84 | 3.18 | 3.23 | 3.16 | 2.91 | 2.78 | 2.75 | 2.77 | 2.89 |
| log D (D in m²/s) Cs | -13.03 0.27 | -12.88 0.16 | -12.72 0.14 | -12.42 0.12 | -12.26 0.11 | -12.24 0.16 | -12.05 0.09 | -11.78 0.12 | -11.92 0.09 | -11.68 0.07 | -12.38 0.08 |
| Rb | -12.76 0.24 | -12.81 0.17 | -12.75 0.17 | -12.24 0.21 | -12.12 0.11 | -11.99 0.10 | -11.90 0.12 | -11.60 0.08 | -11.54 0.09 | -11.50 0.08 | -12.01 0.08 |
| Ba | -12.94 0.18 | -12.53 0.15 | -12.68 0.23 | -12.06 0.13 | -11.92 0.14 | -11.91 0.14 | -11.81 0.09 | -11.60 0.14 | -11.46 0.11 | -11.51 0.08 | -11.92 0.09 |
| Sr | -13.00 0.13 | -12.59 0.16 | -12.64 0.27 | -12.05 0.11 | -12.13 0.11 | -11.93 0.08 | -11.66 0.09 | -11.67 0.13 | -11.60 0.10 | -11.60 0.07 | -11.82 0.10 |
| Co | -13.06 0.19 | -12.86 0.16 | -12.64 0.14 | -12.38 0.16 | -12.24 0.10 | -12.06 0.13 | -12.13 0.14 | -11.87 0.12 | -11.67 0.10 | -11.71 0.07 | -12.14 0.12 |
| Eu | -13.05 0.24 | -12.84 0.16 | -12.63 0.22 | -12.44 0.14 | -12.55 0.14 | -12.34 0.10 | -12.28 0.12 | -12.17 0.19 | -11.89 0.15 | -11.79 0.07 | -12.37 0.11 |
| Ta | -12.96 0.30 | -12.99 0.18 | -12.65 0.21 | -12.55 0.23 | -12.32 0.16 | -12.45 0.16 | -12.32 0.17 | -12.28 0.18 | -12.00 0.16 | -12.03 0.25 | -12.36 0.18 |
| V | -13.09 0.12 | -13.10 0.17 | -12.78 0.20 | -12.67 0.12 | -12.64 0.10 | -12.51 0.12 | -12.54 0.12 | -12.28 0.09 | -12.15 0.08 | -12.14 0.09 | -12.49 0.12 |
| Cr | -12.86 0.20 | -12.85 0.19 | -12.71 0.22 | -12.42 0.14 | -12.53 0.17 | -12.21 0.12 | -12.27 0.19 | -12.11 0.12 | -12.00 0.19 | -12.10 0.15 | -12.59 0.13 |
| Th | -13.08 0.24 | -12.83 0.16 | -13.00 0.19 | -12.60 0.17 | -12.30 0.13 | -12.35 0.17 | -12.40 0.12 | -12.04 0.11 | -11.84 0.10 | -11.73 0.14 | -12.21 0.18 |
| U | -13.16 0.19 | -12.92 0.18 | -12.72 0.21 | -12.50 0.13 | -12.49 0.14 | -12.50 0.18 | -12.20 0.16 | -11.91 0.15 | -11.82 0.14 | -11.85 0.11 | -12.00 0.13 |
| Hf | -13.10 0.30 | -12.79 0.23 | -12.69 0.24 | -12.66 0.19 | -12.65 0.13 | -12.51 0.21 | -12.42 0.20 | -12.12 0.11 | -12.03 0.22 | -12.06 0.13 | -12.08 0.18 |
| Sn | -12.02 0.17 | -12.57 0.15 | -12.38 0.24 | -11.48 0.13 | -11.64 0.08 | -11.83 0.16 | -11.37 0.13 | -11.40 0.15 | -11.41 0.16 | -11.46 0.11 | -11.62 0.15 |

**Table 4**

| | Lt$_{58}$ | | | | | Tr$_{62}$ | | | | | Tr$_{66}$ | | | | | Rh$_{70}$ | | | | |
|---|---|---|---|---|---|---|---|---|---|---|---|---|---|---|---|---|---|---|---|---|
| | *a* | | *b* | | R$^2$ | *a* | | *b* | | R$^2$ | *a* | | *b* | | R$^2$ | *a* | | *b* | | R$^2$ |
| **Cs** | 0.48 | *0.05* | -12.46 | *0.07* | 0.91 | 0.56 | *0.05* | -12.75 | *0.07* | 0.93 | 0.58 | 0.06 | -12.89 | 0.09 | 0.90 | 0.62 | *0.06* | -13.05 | *0.07* | 0.87 |
| **Rb** | 0.58 | *0.06* | -12.45 | *0.07* | 0.92 | 0.65 | *0.05* | -12.71 | *0.07* | 0.95 | 0.67 | 0.05 | -12.82 | 0.07 | 0.95 | 0.72 | *0.04* | -12.97 | *0.05* | 0.95 |
| **Ba** | 0.65 | *0.07* | -12.75 | *0.10* | 0.89 | 0.70 | *0.07* | -12.89 | *0.10* | 0.91 | 0.70 | 0.07 | -12.89 | 0.10 | 0.91 | 0.70 | *0.07* | -12.87 | *0.09* | 0.91 |
| **Sr** | 0.61 | *0.07* | -12.67 | *0.09* | 0.91 | 0.66 | *0.05* | -12.84 | *0.07* | 0.94 | 0.69 | 0.06 | -12.90 | 0.08 | 0.94 | 0.70 | *0.07* | -12.90 | *0.09* | 0.93 |
| **Co** | 0.59 | *0.08* | -12.60 | *0.11* | 0.85 | 0.61 | *0.08* | -12.77 | *0.10* | 0.88 | 0.64 | 0.07 | -12.91 | 0.10 | 0.89 | 0.64 | *0.07* | -13.02 | *0.10* | 0.89 |
| **Eu** | 0.57 | *0.06* | -12.69 | *0.08* | 0.91 | 0.54 | *0.07* | -12.80 | *0.09* | 0.88 | 0.53 | 0.07 | -12.90 | 0.10 | 0.85 | 0.52 | *0.07* | -13.01 | *0.10* | 0.84 |
| **Ta** | 0.43 | *0.06* | -12.30 | *0.08* | 0.85 | 0.46 | *0.07* | -12.61 | *0.09* | 0.84 | 0.50 | 0.07 | -12.86 | 0.10 | 0.85 | 0.47 | *0.07* | -12.98 | *0.09* | 0.83 |
| **V** | 0.52 | *0.05* | -12.64 | *0.06* | 0.93 | 0.51 | *0.04* | -12.84 | *0.05* | 0.95 | 0.51 | 0.05 | -13.00 | 0.07 | 0.91 | 0.46 | *0.06* | -13.13 | *0.08* | 0.88 |
| **Cr** | 0.46 | *0.04* | -12.53 | *0.05* | 0.94 | 0.46 | *0.05* | -12.68 | *0.06* | 0.91 | 0.47 | 0.06 | -12.83 | 0.08 | 0.88 | 0.41 | *0.04* | -12.92 | *0.06* | 0.78 |
| **Th** | 0.58 | *0.05* | -12.41 | *0.06* | 0.95 | 0.56 | *0.05* | -12.66 | *0.07* | 0.93 | 0.64 | 0.06 | -12.98 | 0.08 | 0.93 | 0.64 | *0.08* | -13.15 | *0.11* | 0.88 |
| **U** | 0.65 | *0.05* | -12.45 | *0.07* | 0.94 | 0.67 | *0.06* | -12.81 | *0.08* | 0.93 | 0.68 | 0.05 | -13.02 | 0.07 | 0.95 | 0.66 | *0.06* | -13.16 | *0.08* | 0.94 |
| **Hf** | 0.55 | *0.04* | -12.77 | *0.06* | 0.94 | 0.49 | *0.05* | -12.84 | *0.07* | 0.90 | 0.50 | 0.06 | -12.97 | 0.08 | 0.90 | 0.50 | *0.07* | -13.06 | *0.09* | 0.86 |
| **Sn** | 0.55 | *0.05* | -11.83 | *0.07* | 0.93 | 0.54 | *0.07* | -12.07 | *0.10* | 0.87 | 0.58 | 0.09 | -12.31 | 0.13 | 0.81 | 0.55 | *0.11* | -12.40 | *0.14* | 0.77 |

**Table 5**

|    | $D_i/D_{SiO2}$ | $C_f$ (ppm) | $K_{Sho/Rhy}$ |
|----|----------------|-------------|---------------|
| La | 3.5 | 30 | 3.3 |
| Ce | 3.5 | 55 | 3.3 |
| Pr | 3.5 | 6 | 3.0 |
| Nd | 3.0 | 20 | 2.8 |
| Sm | 3.0 | 3.5 | 2.7 |
| Gd | 3.0 | 3.0 | 2.8 |
| Tb | 2.5 | 0.3 | 2.1 |
| Dy | 2.5 | 1.8 | 2.1 |
| Ho | 2.5 | 0.3 | 1.9 |
| Er | 2.5 | 0.7 | 1.7 |
| Tm | 2.0 | 0.1 | 1.6 |
| Yb | 2.0 | 0.7 | 1.6 |
| Lu | 2.0 | 0.1 | 1.6 |
| Y  | 3.0 | 8 | 1.8 |
| Zr | 3.0 | 10 | 1.1 |
| Nb | 2.0 | 5 | 1.5 |
| Pb | 10 | 4 | 1.7 |
| Sr | 10 | 250 | 2.5 |
| Eu | 3.0 | 0.45 | 2.8 |

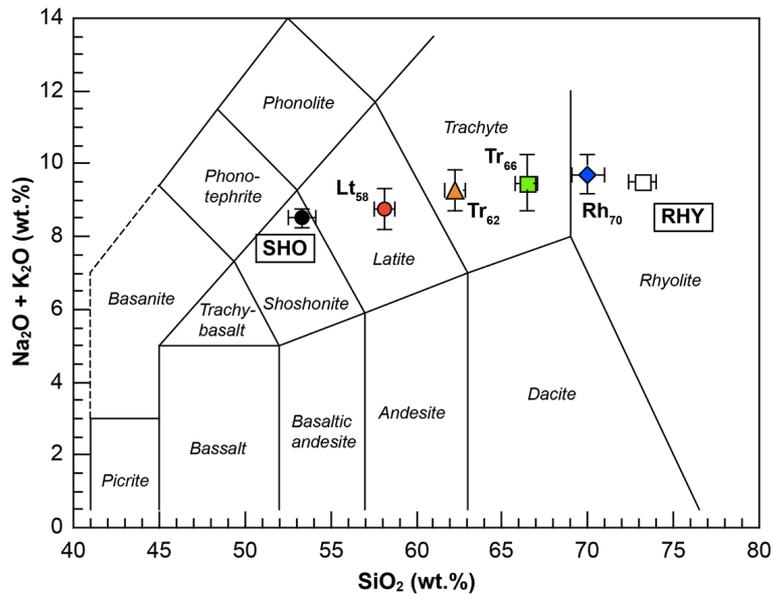

Figure 1

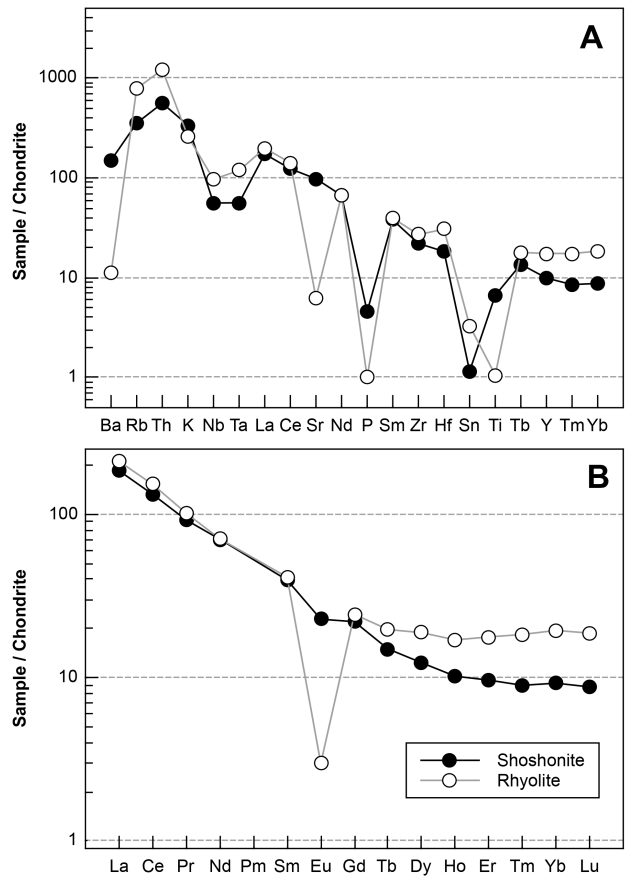

Figure 2

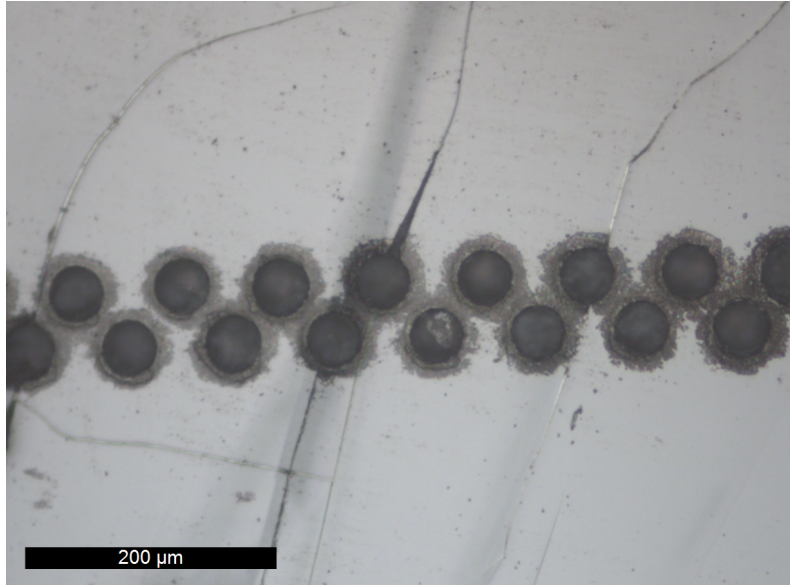

Figure 3

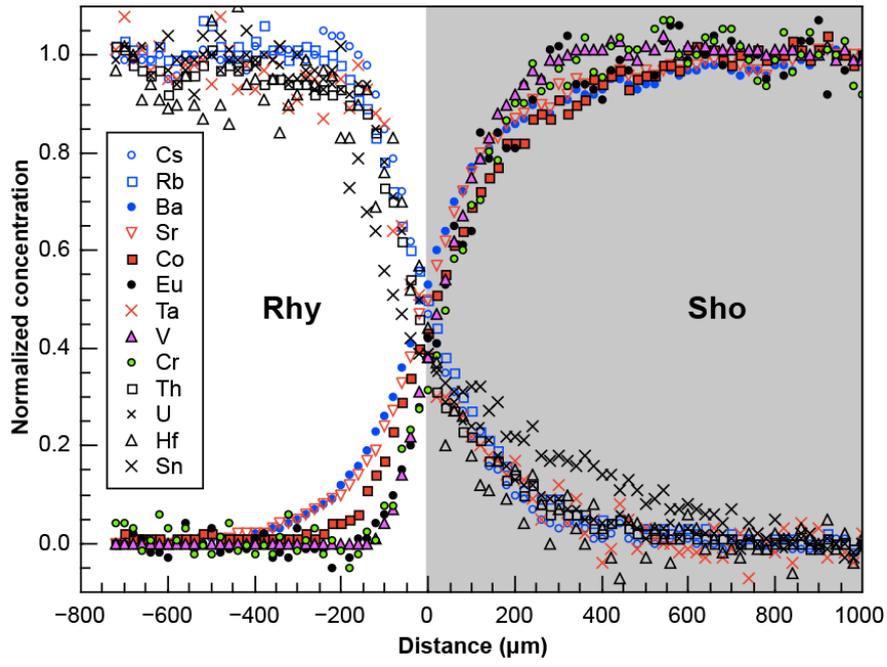

Figure 4

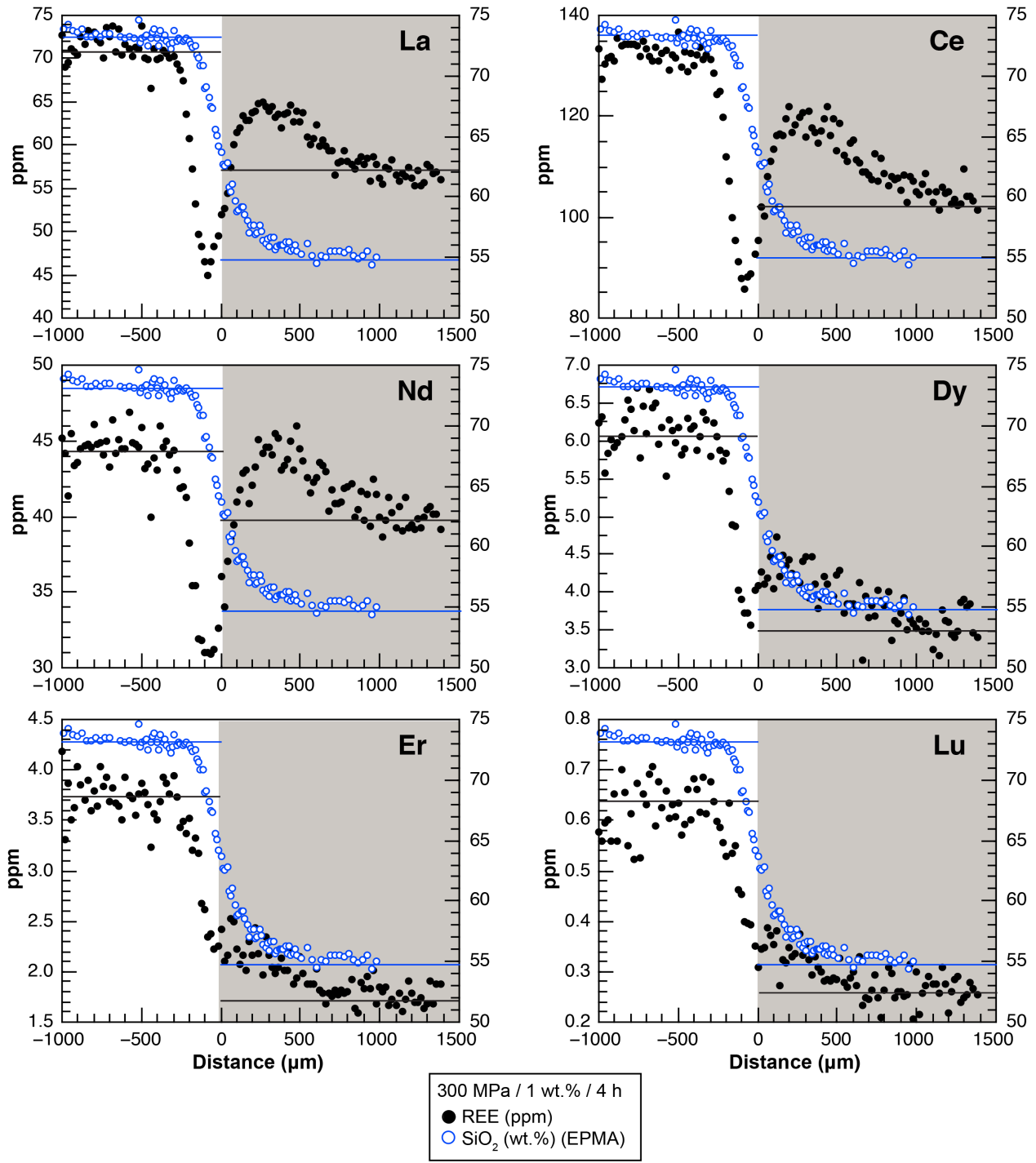

Figure 5

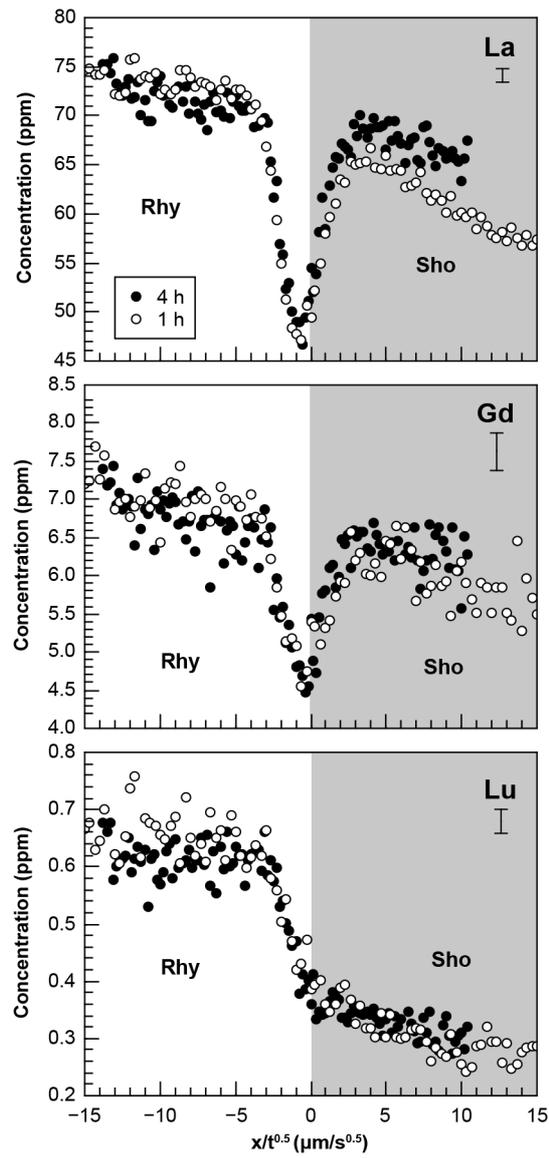

Figure 6

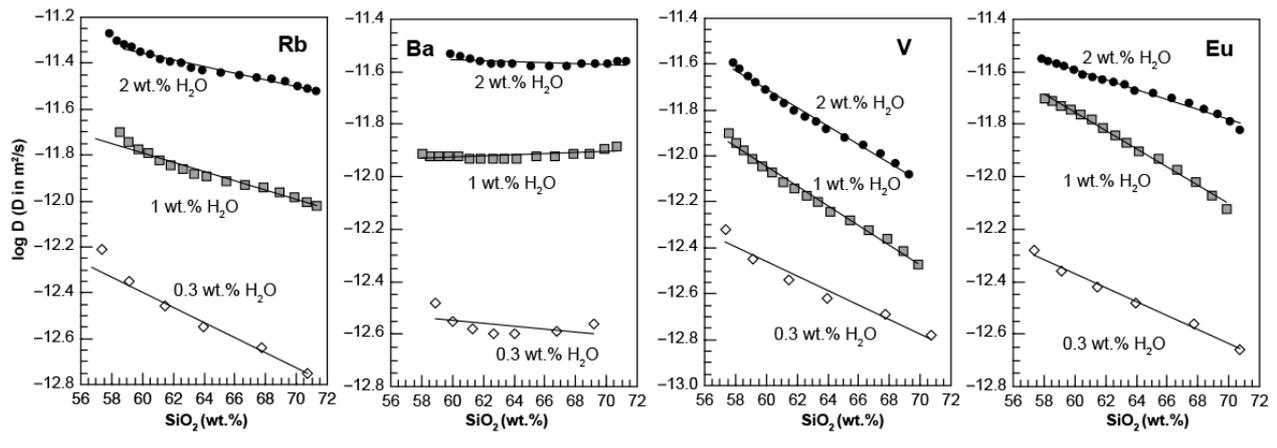

Figure 7

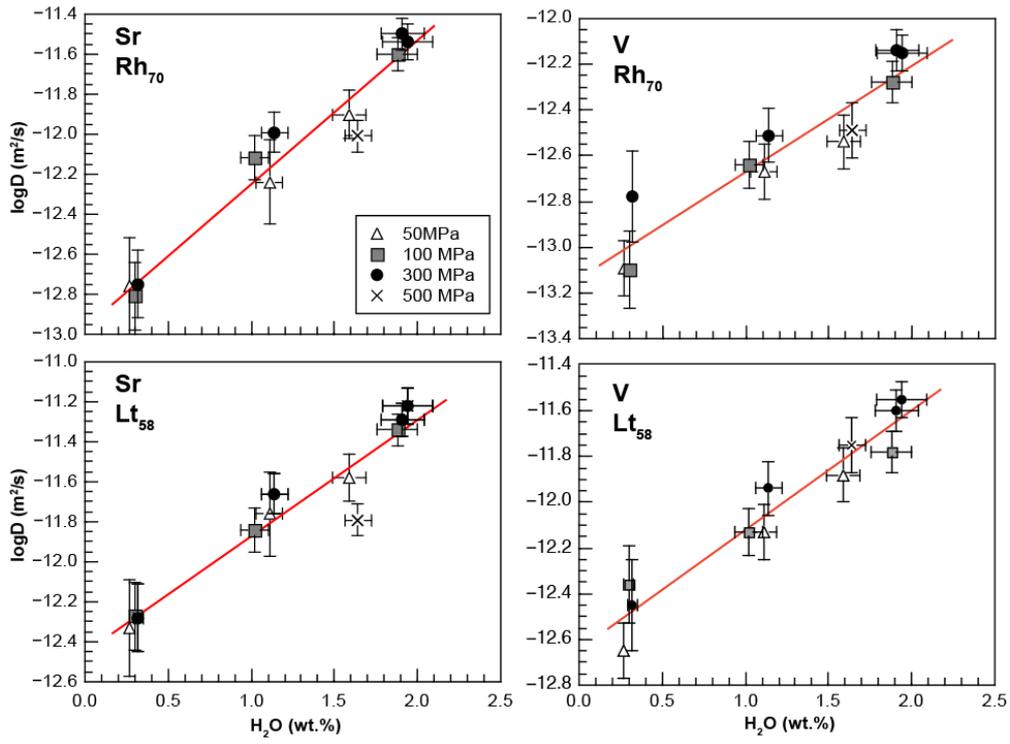

Figure 8

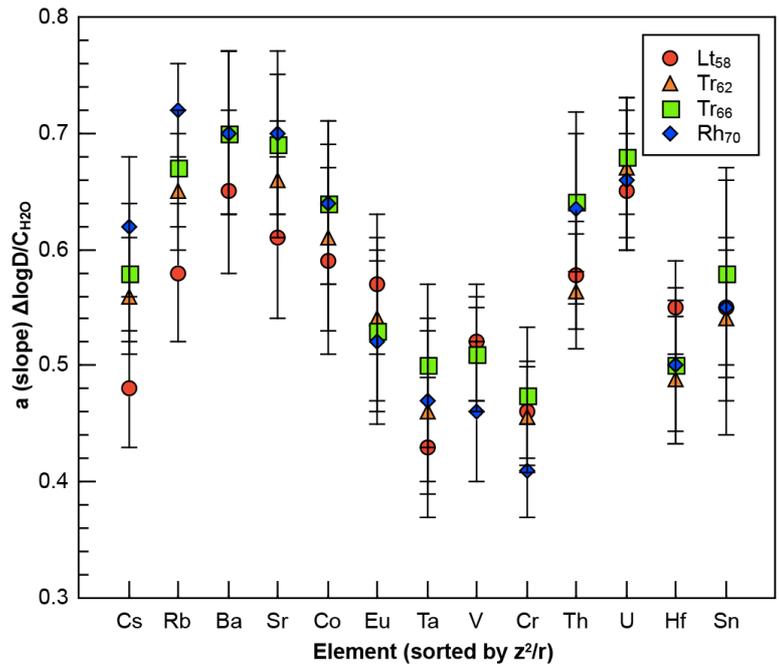

Figure 9

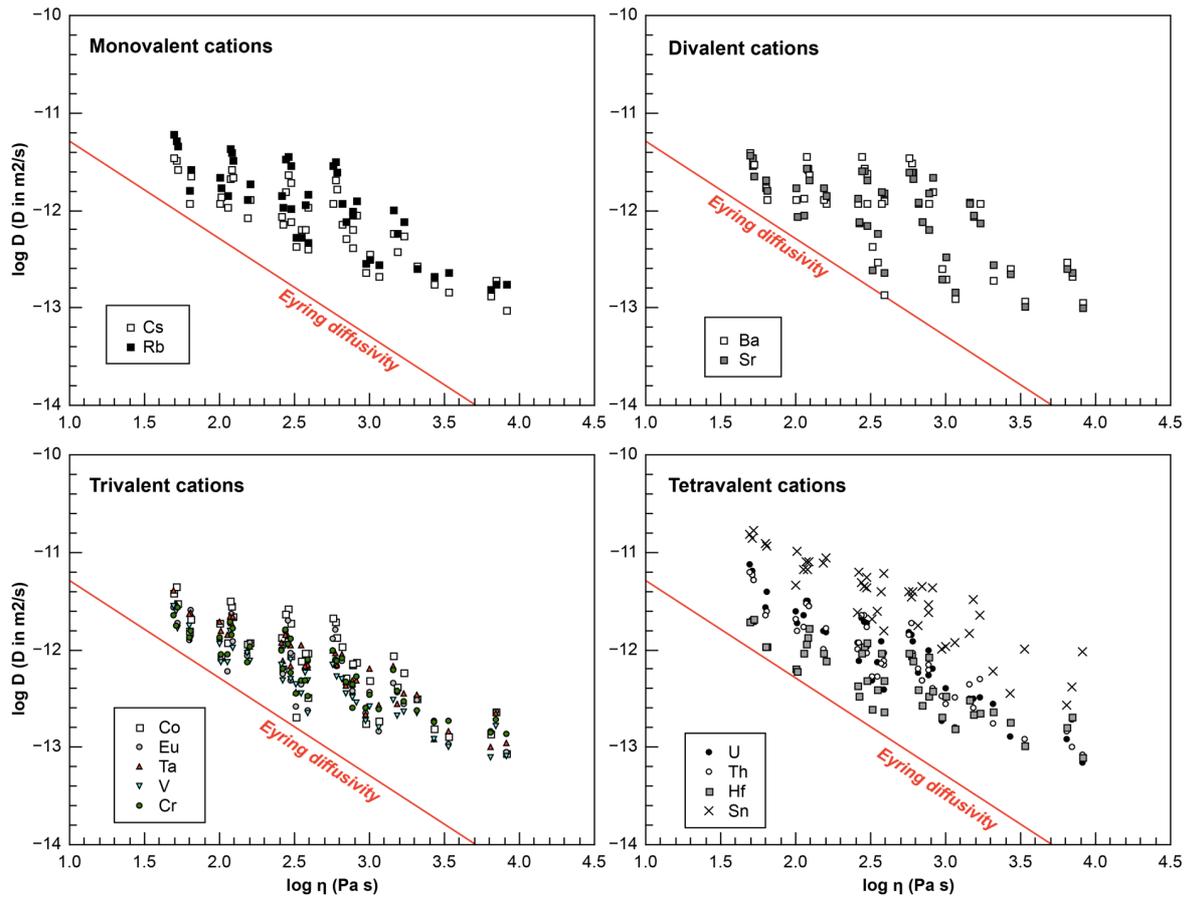

Figure 10

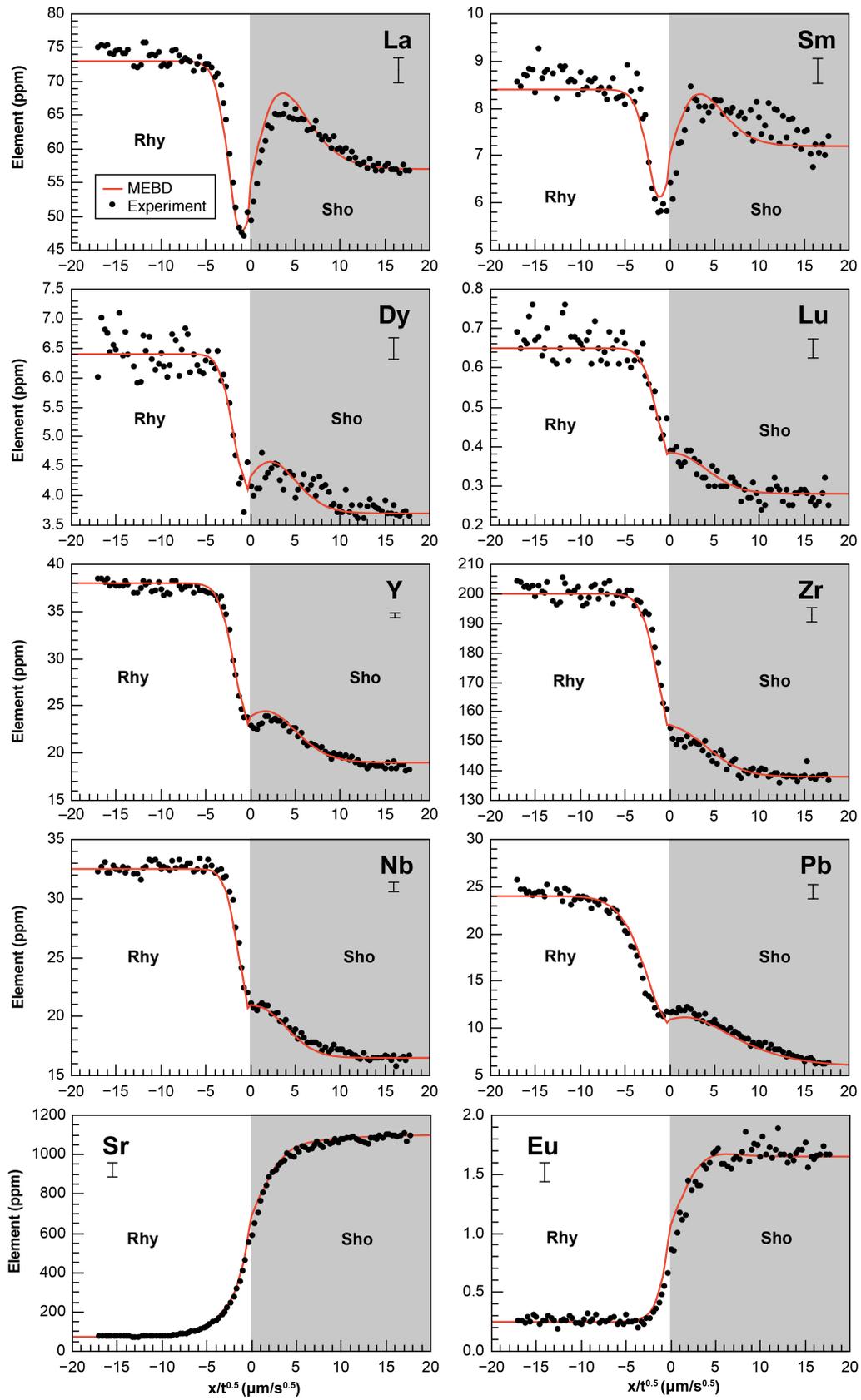

Figure 11

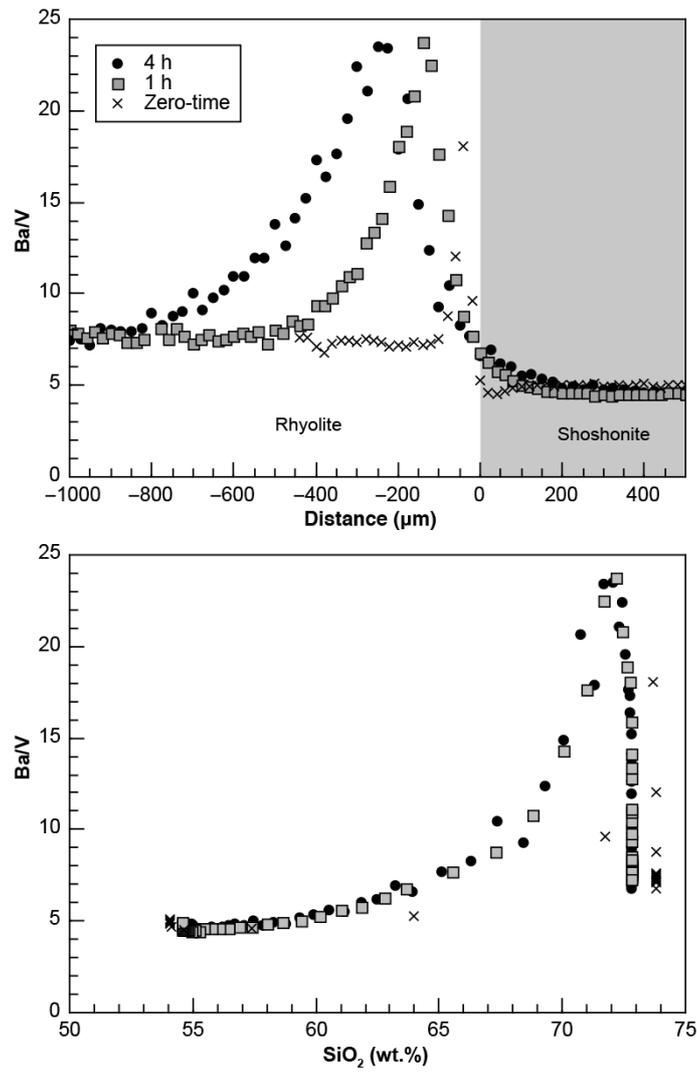

Figure 12